\documentclass[prd,superscriptaddress,twocolumn,letterpaper,%
nofootinbib,showpacs,preprintnumbers,amsmath,amssymb]{revtex4}
\usepackage{bm,curves}
\newcommand{\del}{\partial}
\newcommand{\run}[1]{\widetilde{\alpha}_{#1}}
\newcommand{\hsp}[1]{\hspace*{#1 mm}}

\begin{document}

\title{Matching functions for heavy particles}
\author{S.D. Bass}
\affiliation{High Energy Physics Group,
Institute for Experimental Physics and
Institute for Theoretical Physics, Universit\"at Innsbruck,
Technikerstrasse 25, A 6020 Innsbruck, Austria}
\author{R.J. Crewther}
\affiliation{Department of Physics and Mathematical Physics and \linebreak
Special Research Centre for the Subatomic Structure of Matter (CSSM),
University of Adelaide, SA 5005, Australia}
\affiliation{Centre for Particle Theory, Department of Mathematical
Sciences, University of Durham, South Rd, Durham DH1 3LE, United
Kingdom}
\author{F.M. Steffens}
\affiliation{Instituto de Fisica Teorica,
Rua Pamplona 145, 01405-900 Sao Paulo - SP, Brazil}
\affiliation{Mackenzie University - FCBEE, Rua da Consolacao 930,
01302-907, Sao Paulo - SP, Brazil}
\author{A.W. Thomas}
\affiliation{Department of Physics and Mathematical Physics and \linebreak
Special Research Centre for the Subatomic Structure of Matter (CSSM),
University of Adelaide, SA 5005, Australia}

\begin{abstract}
We introduce \textit{matching functions} as a means of summing 
heavy-quark logarithms to any order.  Our analysis is based on 
Witten's approach, where heavy quarks are decoupled one at a
time in a mass-independent renormalization scheme.  The outcome is a
generalization of the matching conditions of Bernreuther and Wetzel:  
we show how to derive closed formulas for summed logarithms to any order, 
and present explicit expressions for leading order (LO) and 
next-to-leading order (NLO) contributions.  The decoupling of
heavy particles in theories lacking asymptotic freedom is also 
considered.
\end{abstract}
\preprint{ADP-02-100/T538}
\maketitle

\section{Introduction}\label{I}

Decoupling a heavy quark when the renormalization scheme is mass 
independent was originally discussed by Witten \cite{witten}.  
He showed that the results can be elegantly expressed in terms of 
a renormalization group (RG) invariant running coupling 
$\run{h}$ associated with the mass $m_h$ of the heavy quark $h$.  
Subsequently, Bernreuther and Wetzel \cite{BW,wetzel,B1,B2} 
proposed a systematic method for dealing with the matching problem, 
i.e.\ the lack of explicit decoupling in mass independent schemes.  
They applied the Appelquist-Carrazone decoupling theorem 
\cite{AC} to the gluon coupling $\alpha_Q^{\rm MO}$ in the momentum
subtraction (MO) scheme, i.e.\ renormalized at space-like momentum $Q$
\begin{equation}
\alpha^{\mbox{\tiny MO}}_Q\bigr|_{\mathrm{with}\ h}
   = \alpha^{\mbox{\tiny MO}}_Q\bigr|_{\mathrm{no}\ h}\,
          +\,  O(m^{-1}_h)
\label{aa1}
\end{equation}
and compared calculations of $\alpha^{\mbox{\tiny MO}}_Q$ in the full
$F=f+1$ and effective $f$ flavor $\overline{\mbox{\small MS}}$ 
(modified minimal subtraction) theories. When $O(m^{-1}_h)$ terms 
are neglected, the strong coupling $\alpha_F = g_F^2/(4\pi)$ for the 
$F$-flavor $\overline{\mbox{\small MS}}$ theory is calculable as a 
power series in its $f$-flavor counterpart $\alpha_f$ and logarithms of 
$m_h$. Results for the first few loops of perturbation theory appear in 
the literature \cite{BW,wetzel,B1,B2,larin,CKS1,rodrigo}. Bernreuther has 
constructed a similar matching procedure to deal with the effects of 
mass renormalization \cite{B1}.

This paper arises from the observation \cite{BW,B1} that the RG relates
coefficients of perturbative mass logarithms $\sim \alpha_F^r\ln^s m_h$
in matching relations.  This suggests that we seek an analogy with 
the behavior of Green's functions at large momenta $q$, where in general 
\cite{GML,IZ}, each perturbative order in the Gell-Mann--Low function 
$\Psi(x)$ or the Callan-Symanzik $\beta,\gamma,\delta$ functions 
determines all coefficients to logarithmic order $\kappa$, i.e.\ 
all coefficients of momentum logarithms 
$\{\alpha_F^{r+\kappa}\ln^r q,\ r = 0,1,2,\ldots\}$.

We set up the formalism for mass logarithms by introducing 
\textit{matching functions} ${\cal F}(\run{h})$ and ${\cal G}(\run{h})$ 
associated with coupling constant and mass renormalization.  The coupling
constant and masses are matched to all orders, with heavy quarks (in 
this article) decoupled one by one ($F=f+1$).  {}For coupling 
constant renormalization, we derive in Sec.~\ref{mf} the following
key equations, 
\begin{align}
\ln\frac{\bar{m}_h}{\bar{\mu}}
&= \int_{\alpha_F}^{\run{h}}\hsp{-2.5}dx\,\frac{1}{\beta_F(x)} 
\label{c8} \\[1mm]
\ln\frac{\bar{m}_h}{\bar{\mu}}
&= \int_{\alpha_f}^{\run{h}}\hsp{-2.5}dx\,\frac{1}{\beta_f(x)} 
       + {\cal F}_{F\to f}(\run{h}) 
\label{c9} \end{align}
where ${\bar{m}_h}$ is Witten's renormalization group invariant 
heavy-quark mass \cite{witten}, $\beta_f(x)$ is the 
$\overline{\mbox{\small MS}}$ $\beta$-function for the $f$-flavor 
theory, and 
\begin{equation}
\bar{\mu} 
= \mu_\mathrm{dim}\sqrt{4\pi}e^{-\gamma/2}\ ,\ \ \gamma = 0.5772 \ldots
\end{equation}
is the $\overline{\mbox{\small MS}}$ scale derived from the scale 
$\mu_\mathrm{dim}$ used to define dimensional regularization and
renormalization.  The matching function ${\cal F}_{F\to f}$ is a 
series in $\run{h}$ whose coefficients can be determined perturbatively 
by comparison with (\ref{aa1}).  The desired matching relation between 
$\alpha_F$ and $\alpha_f$ is the result of eliminating 
$\run{h}$ from Eqs.~(\ref{c8}) and (\ref{c9}):
\begin{equation}
\alpha_F = \alpha_F\bigl(\alpha_f,\ln(\bar{m}_h/\bar{\mu})\bigr) .
\label{d0}
\end{equation}
Similar conditions for mass matching are presented in Sec.~\ref{mmf}.

The role played by ${\cal F}$, ${\cal G}$ and the $\beta,\gamma,\delta$ 
functions in matching conditions is just like that of the 
$\beta,\gamma,\delta$ functions for large-momentum logarithms.  Each
order of perturbation theory for these functions determines the 
coefficients of a new tower of mass logarithms:  leading order (LO), 
next-to-leading order (NLO), next-to-next-to-leading order (NNLO), and 
so on.  Both ${\cal F}(\run{h})$ and ${\cal G}(\run{h})$ vanish for LO 
and NLO, but then there are contributions from successive terms in their 
power series in $\run{h}$, starting with NNLO for ${\cal F}(\run{h})$
and NNNLO for ${\cal G}(\run{h})$.

We find, as for large-$q$ logarithms, that results for coupling constants, 
running couplings and light masses can be most elegantly expressed as
closed expressions or generating functions for summed towers of mass 
logarithms.  Examples are the decoupling formulas for $\alpha_{f+1}$ and 
$\run{h}$ quoted previously by us \cite{BCST}, which we derive in  
Sec.~\ref{ce}.  Almost all of our results are for quantum chromodynamics 
(QCD) with three colors, but the technique can be applied to any 
renormalizable theory.

Sections \ref{wm} and \ref{mcc} are brief summaries of Witten's treatment 
of heavy-quark decoupling in QCD and the matching procedure of Bernreuther 
and Wetzel for coupling constants.  This lays the foundation for the RG 
analysis in Sec.~\ref{mf}, from which we are led to construct the matching 
function ${\cal F}$ for coupling constant renormalization.  Perturbation
theory for ${\cal F}$ is considered in Sec.~\ref{pt}, with the result that
the first non-zero term (NNLO) in ${\cal F}$ is obtained.  In Sec.~\ref{ce}, 
we show that Eqs.~(\ref{c8}) and (\ref{c9}) lead directly to closed 
expressions for heavy-quark logarithms to a given logarithmic order, and
present explicit NLO expressions.  Section~\ref{mmf} is an extension of
our RG analysis to deal with the matching problem for mass renormalization.  
It is here that we introduce the mass-matching function ${\cal G}$.  
In Sec.~\ref{asd}, we decouple more than one heavy quark sequentially, 
for example  
$\ln (m_t / {\bar \mu}) \gg \ln (m_b / {\bar \mu}) \rightarrow \infty$,
and derive the NLO closed formula for coupling constant 
renormalization in this limit.

Section~\ref{nonAF} suggests that the consistency of theories lacking 
asymptotic freedom, such as quantum electrodynamics (QED) with heavy 
leptons, be tested by imposing the physical requirement that all heavy 
particles decouple in the infinite-mass limit.  Both ultra-violet and 
infra-red stable fixed points enter the analysis.

Other applications of our technique are discussed in the concluding 
Sec.~\ref{c}.

\section{Witten's Method}\label{wm}
This section summarizes some key points of Witten's procedure
\cite{witten,BCST}.

By convention, the same $\overline{\mbox{\small MS}}$ scale $\bar{\mu}$ 
is used for the initial $F$-flavor and all residual $f$-flavor theories.  
Whenever heavy quarks (masses $m_h$) are decoupled,
\[ F\to f\ \mbox{flavors},\,\ m_h \to \infty  \]
all parameters of the \emph{residual} $f$-flavor theory are 
held fixed: the scale $\bar{\mu}$, all momenta $\mathbf{p}$,
the coupling $\alpha_f$, and all light-quark masses $m_{lf}$.  
In any order of perturbation theory, amplitudes
\[ {\cal A}_F =
{\cal A}_F\bigl(\mathbf{p}, \bar{\mu}, \alpha_F, m_{lF}, m_h\bigr)
\]
are power series in $\smash{m_h^{-1}}$ with each power modified by a
polynomial in $\ln(m_h/\bar{\mu})$.  We will consider the leading power 
$\widetilde{\cal A}_F$:
\begin{equation}
{\cal A}_F = \widetilde{\cal A}_F\{1 + O(m_h^{-1})\}.
\end{equation}
The notation $O(m_h^{-1})$ refers to any sub-leading power, including 
its logarithmic modifications.

Logarithms in $\widetilde{\cal A}_F$ for $m_h\sim\infty$ are generated
by 1PI (one-particle irreducible) subgraphs with at least one 
heavy-quark propagator and with degree of divergence at least logarithmic.  
It is as if all contributing 1PI parts were shrunk to a point.  All 
$F$-flavor amplitudes $\widetilde{\cal A}_F$ tend to amplitudes 
${\cal A}_f$ of the residual $f$-flavor theory, apart from 
$m_h$-dependent renormalizations of the coupling constant, light masses, 
and amplitudes \cite{AC}:
\begin{align}
\widetilde{\cal A}_F\bigl(\mathbf{p}&, \bar{\mu},\alpha_F,m_{lF},m_h\bigr)
  \nonumber \\
&=\, \sum_{{\cal A}'}{\cal Z}_{{\cal AA}'}(\alpha_F, m_h/\bar{\mu})
{\cal A}'_f\bigl({\bf p},\bar{\mu},\alpha_f,m_{lf}\bigr)
\label{a} \\
\alpha_f =\ &\alpha_f(\alpha_F, m_h/\bar{\mu}) \hsp{1},\hsp{1}
m_{lf} = m_{lF}D(\alpha_F, m_h/\bar{\mu}).
\label{a1}\end{align}
For practical applications, Eq.~(\ref{a1}) has to be inverted, so that
$\alpha_f$ and $m_{lf}$ become the dependent variables instead of 
$\alpha_F$ and $m_{lF}$. That is because we hold $\alpha_f$ and 
$m_{lf}$ fixed as $m_h \to \infty$.

For any number of flavors $f$ (including $F$), let
\begin{equation}
{\cal D}_f\, =\, \bar{\mu}\frac{\del\ }{\del\bar{\mu}}
            + \beta_f(\alpha_f)\frac{\del\ \ }{\del\alpha_f}
        + \delta_f(\alpha_f)\sum_{k=1}^f {m_k}_{\hsp{-0.2}f}
               \frac{\del\hsp{4.6}}{\del {m_k}_{\hsp{-0.2}f}}
\label{b7}
\end{equation}
be the corresponding Callan-Symanzik operator.  Since ${\cal A}_F$ 
satisfies an $F$-flavor improved Callan-Symanzik equation 
\cite{hooft}, so also does its leading power:
\begin{equation}
\bigl\{{\cal D}_F + \gamma_F(\alpha_F)\bigr\}\widetilde{\cal A}_F\, =\, 0 .
\label{d}
\end{equation}
In general, both  $\gamma_F$ and
${\cal Z} = \bigl({\cal Z}_{{\cal AA}'}\bigr)$ are matrices.

If we substitute (\ref{a}) in (\ref{d}) and change variables,
\begin{equation}
{\cal D}_F\, =\, \bar{\mu}\frac{\del\ }{\del\bar{\mu}}
   + \bigl({\cal D}_F\alpha_f\bigr)\frac{\del\ \ }{\del\alpha_f}
   + \sum_{k=1}^f\bigl({\cal D}_F{m_k}_{\hsp{-0.2}f}\bigr)
              \frac{\del\hsp{4.6}}{\del {m_k}_{\hsp{-0.2}f}}
\end{equation}
the result is an improved Callan-Symanzik equation for each residual
amplitude,
\begin{equation}
\bigl\{{\cal D}_f + \gamma_f(\alpha_f)\bigr\}{\cal A}_f\, =\, 0
\end{equation}
where the functions \cite{witten,BW}
\begin{eqnarray}
\beta_f(\alpha_f) &=& {\cal D}_F\alpha_f
\label{e1} \\
\delta_f(\alpha_f) &=& {\cal D}_F\ln m_{lf}
\label{eq:e2} \\
\gamma_f(\alpha_f) &=&
     {\cal Z}^{-1}\bigl(\gamma_F(\alpha_F) + {\cal D}_F\bigr){\cal Z}
\label{e3}
\end{eqnarray}
depend \emph{solely} on $\alpha_f$.  The absence of $m_l$ dependence
in the renormalization factors in (\ref{a}) and (\ref{a1}) ensures
mass-independent renormalization for the residual theory.

While these equations hold for any $f<F$, their solutions can be readily
formulated in terms of running couplings only when the heavy quarks 
are decoupled one at a time.  Indeed, Witten's running coupling
\begin{equation}
\run{h} = \run{h}\bigl(\alpha_F, \ln(m_h/\bar{\mu})\bigr)
\end{equation}
is defined for the case $F = f+1$ where just one quark $h$ is heavy, with 
$\overline{\mbox{\small MS}}_F$ renormalized mass $m_h$. The definition
of $\run{h}$ is formulated implicitly \cite{witten}:
\begin{equation}
\ln(m_h/\bar{\mu})
= \int^{\run{h}}_{\alpha_F}\! dx\,\bigl(1-\delta_F(x)\bigr)
    \!\bigm/\!\beta_F(x) .
\label{f3}
\end{equation}
It satisfies the constraints
\begin{equation}
\run{h}(\alpha_F,0) = \alpha_F \hsp{3},\hsp{3}
\run{h}(\alpha_F,\infty) = 0
\end{equation}
where the latter follows from the asymptotic freedom of the $F$-flavor 
theory ($F \leqslant 16$).  Eqs.~(\ref{b7}), (\ref{e1}) and (\ref{f3}) 
imply that $\run{h}$ is renormalization group (RG) invariant: 
\begin{equation}
{\cal D}_F \run{h} = 0 .
\end{equation}

\section{Matching Coupling Constants}\label{mcc}

Generally, the solutions of Witten's equations depend on 
renormalized parameters $\alpha_F$ and $m_{lF}$ of the original
$F=f+1$ flavor theory, whereas the limit $m_h \to \infty$ is to be taken
with parameters $\alpha_f$ and $m_{lf}$ of the \emph{residual} theory
held fixed. To complete the analysis, it is necessary to derive 
asymptotic series in $\ln(m_h/\bar{\mu})$ which relate the initial 
and residual parameters, i.e.\ to ``match'' $\alpha_F$ and $m_{lF}$ 
with $\alpha_f$ and $m_{lf}$.  As noted in Sec.~\ref{I}, Bernreuther 
and Wetzel \cite{BW,wetzel,B1,B2} have set up a systematic 
procedure for this.  This section is a brief account of their
scheme for the case of coupling-constant matching.

The decoupling formula (\ref{aa1}) works to any order of perturbation 
theory, so the task is to express the leading power of the
RG-invariant gluon coupling $\alpha^{\mbox{\tiny MO}}_Q$ with and 
without the heavy-quark $h$ as perturbative series in $\alpha_F$ and 
$\alpha_f$ respectively. 
Generally this involves gluon and other self-energy insertions and a 
vertex amplitude such as fermion-gluon \cite{BW, wetzel} or ghost-gluon 
\cite{B1}. 

For one-loop contributions \cite{weinberg2}, vertex and propagator 
corrections cancel ($Z_1=Z_2$), so only the gluon self-energy amplitude
\[ \Pi^{ab}_{f\mu\nu} 
   = i\delta^{ab}(g_{\mu\nu}q^2 - q_\mu q_\nu)\Pi_f(\sqrt{-q^2}) \] 
is needed.  In that case, we can make the replacement
\[ \alpha^{\mbox{\tiny MO}}_Q\bigr|_{f\ \mathrm{flavors}}
   \longrightarrow \alpha_f\!\bigm/\!\bigl(1 - \Pi_f(Q)\bigr) \]
in Eq.~(\ref{aa1}), with the result
\begin{align}
\alpha_{f+1}^{\,-1} - \alpha_{f+1}^{\,-1}&\Pi_{f+1}(Q)  \nonumber \\ 
&= \alpha_f^{\,-1} - \alpha_f^{\,-1}\Pi_f(Q)
      +  O(m_h^{-1},\alpha_f^2) . 
\label{a3}\end{align}
Comparing the original and residual theories, we have
\begin{align}
\Pi_{f+1}(Q) &= \alpha_{f+1}\bigl\{\Gamma_{h\mathrm{-loop}} 
             + \Gamma_\mathrm{other}\bigr\} + O(\alpha_{f+1}^2)
\nonumber \\
\Pi_f(Q) &= \alpha_f\Gamma_\mathrm{other} + O(\alpha_f^2)
\label{a4}\end{align}
where
\begin{equation}
\Gamma_{h\mathrm{-loop}} = \frac{1}{\pi}\int^1_0\!\!ds\, s(1\!-\!s)
         \ln\biggl(\frac{m_h^2 + s(1\!-\!s)Q^2}{\bar{\mu}^2}\biggr)
\end{equation}
is the contribution of the heavy-quark loop, and $\Gamma_\mathrm{other}$
represents other one-loop terms. 

The leading power contributed by the heavy-quark loop is
\begin{equation}
\Gamma_{h\mathrm{-loop}} = C_\mathrm{LO}\ln\frac{m_h}{\bar{\mu}}
     + C_\mathrm{NLO} + O\bigl(Q^2/m_h^2\bigr)
\label{a5}\end{equation}
with coefficients for leading and non-leading logarithmic orders given by
\begin{equation}
C_\mathrm{LO} = 1/(3\pi) \hsp{3}\mbox{and}\hsp{3} C_\mathrm{NLO} = 0 .
\label{a5a}
\end{equation} 
The vanishing of the NLO constant term is a well-known characteristic 
of the $\overline{\mbox{\small MS}}$ gluon self-energy \cite{weinberg2}.

Eliminating $\Gamma_\mathrm{other}$ from Eq.~(\ref{a4}) and combining the
result with Eqs.~(\ref{a3}) and (\ref{a5}), we recover the standard 
one-loop matching condition
\begin{equation}
\alpha_{f+1}^{\,-1} - \alpha_f^{\,-1}
 = \frac{1}{3\pi}\ln\frac{m_h}{\bar{\mu}} +  O\bigl(\alpha_f,m_h^{-1}\bigr)
\label{a6}
\end{equation}
or equivalently
\begin{equation}
\alpha_{f+1} = \alpha_f - \frac{\alpha_f^2}{3\pi}\ln\frac{m_h}{\bar{\mu}}
                +  O\bigl(\alpha_f^3,m_h^{-1}\bigr)   .
\label{a7}
\end{equation}

The two-loop analysis is much more complicated, so we simply quote the
result \cite{BW,wetzel,B1,B2}, taking account of a 
subsequent correction \cite{BW,larin,CKS1}.  We find it convenient to 
consider the inverse form where $\alpha_{f+1}$ is written as a series 
in $\alpha_f$.  For the special case of three colors, the result is:
\begin{align}
\alpha_{f+1}
= \alpha_{f} 
   &- \frac{\alpha_{f}^2}{6\pi}\ln\frac{m_h^2}{\bar{\mu}^2}
   + \frac{\alpha_{f}^3}{36\pi^2}\ln^2\frac{m_h^2}{\bar{\mu}^2}
\nonumber \\
   &- \frac{11\alpha_{f}^3}{24\pi^2}\ln\frac{m_h^2}{\bar{\mu}^2}
   - \frac{11\alpha_{f}^3}{72\pi^2}\, +\, O(\alpha_{f}^4) .
\label{e3c}
\end{align}
The first three terms of the right-hand side belong to the leading order
LO, i.e.\ they are proportional to $\alpha_f$ times a power of 
$\{\alpha_f\ln(m_h/\bar{\mu})\}$.  Only the fourth term is NLO; there 
is no $O(\alpha_f^2)$ term independent of $m_h$ because the 
NLO constant in Eq.~(\ref{a5a}) vanishes.  The fifth term is 
$O(\alpha_f^3)$ and $m_h$-independent, so it is the first example of a
NNLO term.  The three-loop result, including the NNNLO constant 
term, is now known \cite{CKS1}.

Now we would like to know what the renormalization group implies for 
matching relations of this type.  Some results for coefficients 
to a given order of perturbation theory already appear in 
\cite{BW,wetzel,B1}.  Consider Eq.\ (3) of Ref.~\cite{BW},
\begin{equation}
\frac{\alpha_f}{\pi} 
= \frac{\alpha_{f+1}}{\pi} 
   + \sum_{k=1}^\infty\Bigl(\frac{\alpha_{f+1}}{\pi}\Bigr)^{k+1}
             C_k\Bigl(\ln\frac{m_h^2}{\mu^2}\Bigr) + O(m_h^{-1})
\label{b1}
\end{equation}
where $C_k$ is a polynomial of degree $k$, as noted below Eq.~(8) of
Ref.~\cite{BW}:
\begin{equation}
C_k\, =\, c_{k,k}\bigl(\ln(m_h^2/\mu^2)\bigr)^k +\, \ldots\, + c_{k,0} .
\label{b2}
\end{equation}
The constants $c_{1,0},\,c_{2,0},\,c_{3,0},\,\ldots$ are the remainders
left when all terms depending on $\ln(m_h^2/\mu^2)$ are subtracted from
the leading-power functions\, $C_1,\,C_2,\,C_3,\ldots$.  Then, if all 
coefficients and RG functions are known to $k-1$ loops, the RG determines 
all $k$-loop coefficients $c_{k,j}$ in $C_k$ \emph{except} 
for $c_{k,0}$. The latter \emph{cannot} be deduced from the RG, to any 
number of loops; rather, $c_{k,0}$ must be calculated explicitly via a 
separate $k$-loop matching calculation.  For example, the NNLO coefficient 
$-11/(72\pi^2)$ in (\ref{e3c}) is just $-c_{2,0}$.

Instead of Eq.~(\ref{b1}), we prefer to consider the inverse relation
\begin{equation}
\alpha_{f+1} = \alpha_f + \sum_{k=1}^\infty \alpha_f^{k+1}
               P_k\Bigl(\ln\frac{m_h}{\bar{\mu}}\Bigr) 
               + O(m_h^{-1})
\end{equation}
because that is what is required in order to take $m_h\to\infty$ with 
$\alpha_f$ held fixed.  The analogue of Eq.~(\ref{b2}) is
\begin{align}
P_k = p_{k,k}\bigl(\ln(m_h/\bar{\mu})\bigr)^k 
          &+ p_{k,k-1}\bigl(\ln(m_h/\bar{\mu})\bigr)^{k-1} 
\nonumber \\
          &+ \ldots + p_{k,0} .
\label{b4}
\end{align}
An analysis in the style of Bernreuther and Wetzel produces
identical conclusions for the remainder constants $p_{k,0}$: 
given $p_{1,0}, p_{2,0},\ldots p_{k-1,0}$, one can use the RG 
to deduce $p_{k,k}, \ldots , p_{k,1}$ but not $p_{k,0}$.

Most practical applications require that terms of the same logarithmic 
order be summed.  This is straightforward for LO logarithms, because 
the LO coefficients $c_{k,k}$ in (\ref{b2}) obey a simple relationship 
\cite{BW}
\begin{equation}
c_{k,k} = \bigl(c_{1,1}\bigr)^{k} 
\end{equation}
which makes the series geometric:
\begin{equation}
\alpha_f\, 
\underset{\mbox{\scriptsize LO}}{=}\, \alpha_{f+1}\!\Bigm/\!\Bigl(1 
    - \frac{\alpha_{f+1}}{3\pi}\ln\frac{m_h}{\bar{\mu}}\Bigr) .
\label{a9}
\end{equation}
This expression is leading order (LO) with respect to powers of 
$\alpha_{f+1}=\alpha_F$ and $\ln(m_h/\bar{\mu})$.  Eq.~(\ref{a9}) 
implies that the term $O(\alpha_f, m_h^{-1})$ in (\ref{a6}) is NLO 
or higher order:
\begin{equation}
\alpha_{f+1}^{\,-1} - \alpha_f^{\,-1}\,
 \underset{\mbox{\scriptsize LO}}{=}\, \frac{1}{3\pi}\ln\frac{m_h}{\bar{\mu}} .
\label{a9a}
\end{equation}
This leads directly to the inverse of (\ref{a9}), 
\begin{equation}
\alpha_{f+1}\, \underset{\mbox{\scriptsize LO}}{=}\, \alpha_f\!\Bigm/\!\Bigl(1 
            + \frac{\alpha_f}{3\pi}\ln\frac{m_h}{\bar{\mu}}\Bigr)
\label{a8}\end{equation}
where now LO refers to powers of $\alpha_f$ and $\ln(m_h/\bar{\mu})$.
Note that the LO coefficients $p_{k,k}$ in Eq.~(\ref{b4}) are given by
\begin{equation}
p_{k,k}\, =\, \bigl(-1\!\bigm/\!3\pi\bigr)^k .
\label{b5}
\end{equation}

Beyond LO, formulas for all the relevant coefficients become
complicated, making order-by-order summation too cumbersome to be practical.
The rest of this paper is concerned with a RG analysis which allows us
to consider matching relations to a given logarithmic order without 
having to expand in perturbative order.

\section{Matching function}\label{mf}

Any RG analysis of decoupling involves ar least \emph{two} renormalization
groups:  one for the initial $F$-flavor theory, and one for each 
$f$-flavor theory produced as a heavy particle decouples.  We append
a flavor subscript to make the distinction, \emph{viz.}\ RG$_F$ or RG$_f$.

A key observation is that any quantity which is RG$_{F}$ invariant
must also be RG$_{f}$ invariant ($f < F$).  For example, Witten's RG$_F$
invariant running coupling $\run{h}$ must satisfy the condition
\begin{equation}
{\cal D}_f\run{h} 
= \biggl(\bar{\mu}\frac{\del\;}{\del\bar{\mu}}
            + \beta\frac{\del\ }{\del\alpha}\biggr)_{\!f}\run{h}
= 0 .
\label{b9}\end{equation}
Generally, the substitution
\begin{equation}
{\cal D}_F\ \ \longrightarrow\ \ {\cal D}_f  
\end{equation}
works when applied to any quantity which survives the limit
$m_h \to \infty$. An example is the formula
\begin{equation}
{\cal D}_6 \alpha_5\, =\, {\cal D}_5\alpha_5\, =\, \beta_5(\alpha_5)  .
\label{c1}\end{equation}
which agrees with the general result (\ref{e1}).
However, \emph{the converse is not generally true}.  For example, the 
top-quark mass $m_t$ is RG$_{f=5}$ invariant, but it is certainly 
\emph{not} RG$_{f=6}$ invariant.  Therefore a study of the RG for the 
original $F$-flavour theory is both necessary and sufficient for the 
full implications of the RG to be understood.

Our starting point is Witten's definition (\ref{f3}) of the invariant 
running coupling $\run{h}$.  Let us regard this as a formula for 
$\ln(m_h/\bar{\mu})$ in terms of $\run{h}$ and $\alpha_F$.  Specifically,
the right-hand side is an integral from $\alpha_F$ to $\run{h}$ involving 
RG$_F$ functions $\beta_F$ and $\delta_F$.  Can a similar formula be 
constructed from RG$_f$ functions such that this mass logarithm becomes
a function of $\run{h}$ and $\alpha_f$?

If such a formula exists, it must be consistent with the requirements 
of the RG$_F$ group for the \emph{original} theory.  However, mass
renormalization produces an unwelcome $F$ dependence in equations 
such as
\begin{equation}
{\cal D}_F\ln(m_h/\bar{\mu}) = \delta_F(\alpha_F) - 1 .
\label{c1a}
\end{equation}
So let us amend the proposal: instead of the 
$\overline{\mbox{\small MS}}_F$ mass $m_h$, consider Witten's
invariant mass%
\footnote{See Eq.~(16) of \cite{witten}.  Similar effective masses have 
been invented for the cases of large momenta \cite{hooft} and light 
quarks \cite{floratos}.  Their RG invariance makes them useful in
phenomenology \cite{witten,floratos} and lattice calculations \cite{CKS2}.}
\begin{equation}
\bar{m}_h 
 = m_h\exp\int^{\run{h}}_{\alpha_F}\!\!dx\, \delta_F(x)/\beta_F(x) .
\label{c5}
\end{equation}
Since $\bar{m}_h$ is RG$_F$ invariant,
\begin{equation} 
{\cal D}_F \bar{m}_h = 0
\end{equation}
replacing $m_h$ by $\bar{m}_h$ in (\ref{c1a}) eliminates the unwanted 
dependence on $\delta_F$:
\begin{equation}
{\cal D}_F\ln(\bar{m}_h/\bar{\mu}) = - 1 .
\label{c5a}
\end{equation}
Notice that the formula (\ref{c8}) for $\ln(\bar{m}_h/\bar{\mu})$ is an
immediate consequence of the definitions (\ref{f3}) and (\ref{c5}) of 
$\run{h}$ and $\bar{m}_h$.  As a check, ${\cal D}_F$ can be applied to the
right-hand side of (\ref{c8}) to give the result $-1$, in agreement with
Eq.~(\ref{c5a}).

Now observe that, because of Eq.~(\ref{e1}), the replacement $F \to f$ 
everywhere on the right-hand side of (\ref{c8}) produces a quantity
which transforms in the same way under the RG$_F$ of the \emph{original} 
theory:
\begin{equation}
{\cal D}_F\int_{\alpha_f}^{\run{h}}\hsp{-2.5}dx\,\frac{1}{\beta_f(x)}  
 = - 1 .
\label{c5b}
\end{equation}
Comparing Eqs.~(\ref{c5a}) and (\ref{c5b}), we see that a RG$_{F}$ 
invariant quantity ${\cal F}$ can be defined as follows:
\begin{equation}
\ln\frac{\bar{m}_h}{\bar{\mu}}\,
=\, \int_{\alpha_f}^{\run{h}}\hsp{-2.5}dx\,
      \frac{1}{\beta_f(x)}\, + {\cal F} \ \ \Longrightarrow\ \
{\cal D}_F {\cal F}\, =\, 0 .
\label{c6}\end{equation}
Since ${\cal F}$ is dimensionless, it can depend on $\run{h}$, 
but RG$_{F}$ invariance forbids dependence on other dimensionless 
variables, such as $\alpha_F$, $\alpha_f$, $\bar{m}_h/m_h$, or 
$m_h/{\bar \mu}$. We call it the
\textsl{matching function}: 
\begin{equation}
{\cal F}\, =\, {\cal F}_{F\to f}(\run{h}) .
\label{c7}\end{equation}
This completes the derivation of the key equations (\ref{c8}) and 
(\ref{c9}):
\begin{align*}
\ln\frac{\bar{m}_h}{\bar{\mu}}\,
&=\, \int_{\alpha_F}^{\run{h}}\hsp{-2.5}dx\,
      \frac{1}{\beta_F(x)}   \\[1mm]
\ln\frac{\bar{m}_h}{\bar{\mu}}\,
&=\, \int_{\alpha_f}^{\run{h}}\hsp{-2.5}dx\,
      \frac{1}{\beta_f(x)}\, + {\cal F}_{F\to f}(\run{h}) .
\end{align*}
When $\run{h}$ is eliminated from these equations, the desired matching 
relation (\ref{d0}) is obtained:
\[
\alpha_F\, =\, \alpha_F\bigl(\alpha_f,\ln(\bar{m}_h/\bar{\mu})\bigr),
\ \ F=f+1 . \]

\section{Perturbation theory for the matching function}\label{pt}

Perturbative matching relations can be used to determine successive
coefficients in the Taylor series of the matching function 
${\cal F}_{F\to f}(\run{h})$.  One needs to calculate the 
mass-independent terms $c_{k,0}$ in (\ref{b2}) or equivalently 
$p_{k,0}$ in (\ref{b4}).  As noted in Sec.~\ref{mcc}, 
these constants cannot be deduced from the RG. This corresponds
to the fact that ${\cal F}_{F\to f}$ cannot be deduced from the 
RG because it is RG$_{F}$ invariant.

To illustrate the procedure, let us deduce the consequences for 
${\cal F}_{f+1 \to f}$ of the perturbative matching relation (\ref{b1}). 
We need the two-loop $\beta$-function for three colors:
\begin{equation}
\beta_f(x) 
 = - \frac{x^2}{6\pi}(33-2f) - \frac{x^3}{12\pi^2}(153-19f) + O(x^4) .
\label{d1}
\end{equation}
Its reciprocal is
\begin{equation}
\bigl\{\beta_f(x)\bigr\}^{-1}\, 
=\, - \frac{6\pi}{33-2f}
      \Bigl(\frac{1}{x^2} - \frac{b_f}{x} + b_f'\Bigr) + O(x)
\label{d2}
\end{equation}
where $b_f$ stands for the constant
\begin{equation}
b_f\, =\, \frac{1}{2\pi}\frac{153-19f}{33-2f}
\label{d3}
\end{equation}
and $b_f'$ is another constant whose precise value is not of concern here.

The expansion (\ref{d2}) inserted into (\ref{c8}) and (\ref{c9}) 
yields the following equations:
\begin{align}
\frac{33-2(f+1)}{6\pi}\ln\frac{\bar{m}_h}{\bar{\mu}}
&= {\run{h}}^{-1} - \alpha_{f+1}^{-1} 
     + b_{f+1} \ln\frac{\run{h}}{\alpha_{f+1} }  \nonumber \\
&\phantom{=\ } + O(\alpha_f^2)
\label{d4}  \\[1mm]
\frac{33-2f}{6\pi}\ln\frac{\bar{m}_h}{\bar{\mu}}
&= {\run{h}}^{-1} - \alpha_f^{-1} 
     + b_f\ln\frac{\run{h}}{\alpha_f} + O(\alpha_f^2)
\nonumber \\
&\phantom{=\ }
    + \frac{33-2f}{6\pi}{\cal F}_{f+1\to f}(\run{h}).
\label{d5} 
\end{align}
Note that the corrections are $O(\alpha_f^2)$: contributions to 
(\ref{d4}) and (\ref{d5}) from the constant term in (\ref{d2}) are
\[ b_{f+1}'\bigl(\run{h} - \alpha_{f+1}\bigr) = O(\alpha_f^2) 
\ \ \mbox{and}\ \ 
b_f'\bigl(\run{h} - \alpha_f\bigr)  = O(\alpha_f^2)   \]
because $\run{h}$, $\alpha_{f+1}$ and $\alpha_f$ differ by $O(\alpha_f^2)$.

The next step is to eliminate $\run{h}$.  This is partially achieved by
subtracting (\ref{d4}) from (\ref{d5}):
\begin{align}
\frac{1}{3\pi}\ln\frac{\bar{m}_h}{\bar{\mu}}
 =\ &\alpha_{f+1}^{-1} - \alpha_f^{-1} + b_f\ln\frac{\run{h}}{\alpha_f}
     - b_{f+1}\ln\frac{\run{h}}{\alpha_{f+1}}
\nonumber \\ 
    &+ \frac{33-2f}{6\pi}{\cal F}_{f+1\to f}(\run{h}) + O(\alpha_f^2) .
\label{d6}
\end{align}
Since $\ln(\run{h}/\alpha_f)$ and $\ln(\run{h}/\alpha_{f+1})$ 
are both $O(\alpha_f)$, Eqs.\ (\ref{a6}) and (\ref{d6}) imply
\[ {\cal F}_{f+1\to f}(\run{h}) = O(\alpha_f)  \]
and so, from Eqs.~(\ref{d4}) and (\ref{d5}), we conclude
\begin{align}
\alpha_{f+1}/\run{h} 
 &= 1 + \frac{\alpha_f}{6\pi}\Bigl( 33-2(f+1) \Bigr) \ln\frac{m_h}{\bar{\mu}}
      + O(\alpha_f^2)
\nonumber \\
\alpha_f/\run{h} 
 &= 1 + \frac{\alpha_f}{6\pi} \Bigl( 33-2f \Bigr) \ln\frac{m_h}{\bar{\mu}}
      + O(\alpha_f^2) .
\label{d9}
\end{align}
The logarithms of these expressions can then be substituted back into
Eq.~(\ref{d6}), with the result
\begin{align}
\frac{1}{3\pi}\ln\frac{\bar{m}_h}{\bar{\mu}}
  =\ &\alpha_{f+1}^{-1} - \alpha_f^{-1} 
    - \frac{19\alpha_f}{12\pi^2}\ln\frac{m_h}{\bar{\mu}}
\nonumber \\ 
    &+ \frac{33-2f}{6\pi}{\cal F}_{f+1\to f}(\run{h}) + O(\alpha_f^2) .
\label{d9a}
\end{align}

The next step is to relate the logarithms of $\bar{m}_h$ and $m_h$.
First substitute the three-color formula
\begin{equation}
\delta_f(x) = - \frac{2x}{\pi} + O(x^2)
\label{g1}
\end{equation}
into the definition (\ref{c5}) of $\bar{m}_h$,
\begin{equation}
\ln\frac{\bar{m}_h}{m_h} 
 = \frac{12}{33-2(f+1)}\ln\frac{\run{h}}{\alpha_{f+1}} + O(\alpha_f^2)
\end{equation}
and then substitute Eq.~(\ref{d9}) for $\alpha_{f+1}/\run{h}$.  The 
result is:
\begin{equation}
\ln\frac{\bar{m}_h}{\bar{\mu}} 
 = \Bigl(1-\frac{2\alpha_f}{\pi}\Bigr)\ln\frac{m_h}{\bar{\mu}} 
      + O(\alpha_f^2) .
\label{g1a}
\end{equation}

Then the logarithm of $\bar{m}_h$ can be eliminated from 
Eqs.~(\ref{d9a}) and (\ref{g1a}):
\begin{align}
\alpha_{f+1}
= \alpha_{f} 
   &- \frac{\alpha_{f}^2}{3\pi}\ln\frac{m_h}{\bar{\mu}}
   + \frac{\alpha_{f}^3}{9\pi^2}\ln^2\frac{m_h}{\bar{\mu}}
   - \frac{11\alpha_{f}^3}{12\pi^2}\ln\frac{m_h}{\bar{\mu}}
\nonumber \\
   &+ \frac{\alpha_{f}^2}{6\pi}(33-2f){\cal F}_{f+1\to f}(\run{h})
   + O(\alpha_{f}^4) .
\label{g1b}
\end{align}
Comparing this with the two-loop matching condition (\ref{e3c}), 
we see that all mass logarithms are correctly reproduced, and that 
the first non-zero term in the matching function can be deduced from 
the constant NNLO term in (\ref{e3c}):
\begin{equation}
{\cal F}_{f+1\to f}(\run{h}) 
 =  - \frac{11}{12\pi(33-2f)}\hsp{0.2}\run{h} + O(\run{h}^2) .
\label{e3d}
\end{equation}

The $O(\run{h}^2)$ term in ${\cal F}$ can be found by substituting 
(\ref{e3d}) back into Eq.~(\ref{c9}) and repeating the above process 
using the three-loop $\beta_f$ and two-loop $\delta_f$ functions.  
The answer follows by comparison with the known three-loop matching 
condition \cite{CKS1}.  That is the limit of current calculations, 
but in principle, this strategy could be pursued to any order, with 
all mass logarithms correctly reproduced.

\section{Closed expressions for heavy-quark logarithms}\label{ce}

The importance of the matching function ${\cal F}$ is that it allows 
us to work to a given logarithmic order without having to sum mass
logarithms order-by-order in perturbation theory.  Indeed, the role
of ${\cal F}$ is essentially the same as that of the RG functions 
$\beta$, $\gamma$ and $\delta$: each term in the series for ${\cal F}$ 
corresponds to a particular logarithmic order.  For LO and NLO,
${\cal F}$ does not contribute, but NNLO requires that the $O(\run{h})$ 
term in (\ref{e3d}) be included, NNNLO requires the $O(\run{h}^2)$ term,
and so on.

To illustrate, let us derive the closed NLO formula for the
matching relation between $\alpha_{f+1}$ and $\alpha_f$ which we 
announced in \cite{BCST}.

As in the previous section, we insert the expansion (\ref{d2}) 
into (\ref{c8}) and (\ref{c9}), but this time we omit the 
NNLO term ${\cal F}_{f+1\to f}$:
\begin{align}
\frac{33-2(f+1)}{6\pi}\ln\frac{\bar{m}_h}{\bar{\mu}}
&\underset{\mbox{\scriptsize NLO}}{=} 
 {\run{h}}^{-1} - \alpha_{f+1}^{-1} + b_{f+1}\ln\frac{\run{h}}{\alpha_{f+1}}  
\label{d5a}\\
\frac{33-2f}{6\pi}\ln\frac{\bar{m}_h}{\bar{\mu}}
&\underset{\mbox{\scriptsize NLO}}{=} 
 {\run{h}}^{-1} - \alpha_f^{-1} + b_f\ln\frac{\run{h}}{\alpha_f} .
\label{d5b} 
\end{align}
The difference of these two equations is
\begin{align}
\frac{1}{3\pi}\ln\frac{\bar{m}_h}{\bar{\mu}}
\underset{\mbox{\scriptsize NLO}}{=}
\alpha_{f+1}^{-1} - \alpha_f^{-1} 
     &- (b_f-b_{f+1})\ln\frac{\alpha_f}{\run{h}}
\nonumber \\
     &- b_{f+1}\ln\frac{\alpha_f}{\alpha_{f+1}} .
\label{d8}
\end{align}
The logarithms on the right-hand side are already NLO, so we can use
the LO parts of (\ref{d5b}) and (\ref{d8}) to approximate their arguments:
\begin{align}
\alpha_f/\run{h} &\underset{\mbox{\scriptsize LO}}{=}
   1 + \frac{\alpha_f}{6\pi}(33-2f)\ln\frac{m_h}{\bar{\mu}}
\label{d9b} \\
\alpha_f/\alpha_{f+1} &\underset{\mbox{\scriptsize LO}}{=}
  1 + \frac{\alpha_f}{3\pi}\ln\frac{m_h}{\bar{\mu}} .
\label{d9c}
\end{align}
The result is a NLO generalisation of (\ref{a8}):
\begin{align}
&\alpha_{f+1} \underset{\mbox{\scriptsize NLO}}{=}
\nonumber \\
&\alpha_f \!\Bigm/\! \Bigl\{  
   1 + \frac{\alpha_f}{3\pi}\ln\frac{\bar{m}_h}{\bar{\mu}}
+ \alpha_f b_{f+1} \ln
 \Bigl(1 + \frac{\alpha_f}{3\pi}\ln\frac{m_h}{\bar{\mu}}\Bigr)
\nonumber \\
&\phantom{\alpha_f}\
+ \alpha_f (b_f - b_{f+1})\ln \Bigl[1 + 
   \frac{\alpha_f}{6\pi}(33-2f)\ln\frac{m_h}{\bar{\mu}}\Bigr]\Bigr\}  .
\label{e0}
\end{align}

If desired, $\ln(\bar{m}_h/\bar{\mu})$ can be eliminated in favor of 
$\ln({m}_h/\bar{\mu})$.  The leading NLO effects of mass
renormalization are due to the one-loop term of $\delta_f$ given by 
Eq.~(\ref{g1}).  When this term is substituted into the definition
(\ref{c5}) of $\bar{m}_h$, keeping all logarithms to this order, we
find an expression
\begin{equation}
\ln\frac{\bar{m}_h}{\bar{\mu}} 
 \underset{\mbox{\scriptsize NLO}}{=} \ln\frac{m_h}{\bar{\mu}} 
    + \frac{12}{31-2f}\Bigl(\ln\frac{\alpha_f}{\alpha_{f+1}}
            - \ln\frac{\alpha_f}{\run{h}}\Bigr) 
\label{e2}\end{equation}
to which Eqs.~(\ref{d9b}) and (\ref{d9c}) can be readily applied:
\begin{align}
\ln\frac{\bar{m}_h}{\bar{\mu}}
&\underset{\mbox{\scriptsize NLO}}{=} \ln\frac{m_h}{\bar{\mu}}
  + \frac{12}{31\!-\!2f}\ln\Bigl(1 
       + \frac{\alpha_f}{3\pi}\ln\frac{m_h}{\bar{\mu}}\Bigr) 
\nonumber \\
&\phantom{\underset{\mbox{\scriptsize NLO}}{=}}\
   - \frac{12}{31\!-\!2f}\ln\Bigl(
      1 + \frac{\alpha_f}{6\pi}(33\!-\!2f)\ln\frac{m_h}{\bar{\mu}}\Bigr) .
\label{y1}
\end{align}
So, by combining (\ref{e0}) and (\ref{y1}), we arrive at a complete NLO
formula for the matching condition:
\begin{align}
\alpha_{f+1}^{-1} &\underset{\mathrm{NLO}}{=}
\alpha_f^{-1} + \frac{1}{3\pi}\ln\frac{m_h}{\bar{\mu}} +
c_f\ln\Bigl[1 + \frac{\alpha_f}{3\pi}\ln\frac{m_h}{\bar{\mu}}\Bigr]
\nonumber \\[1mm]
&\phantom{\underset{\mathrm{NLO}}{=} \alpha_f^{-1}\ }
+ d_f\ln\Bigl[1 + \frac{\alpha_f}{6\pi}(33-2f)
      \ln\frac{m_h}{\bar{\mu}}\Bigr] 
 , \nonumber \\
c_f &=  \frac{142-19f}{2\pi(31\!-\!2f)} \hsp{2},\hsp{2}
d_f =  \frac{57+16f}{2\pi(33\!-\!2f)(31\!-\!2f)} .
\label{y3}
\end{align}

The same equations can also be used to obtain equations for the RG 
invariant $\run{h}$ (also announced in \cite{BCST}). Eqs.~(\ref{d3}),
(\ref{d5b}) and (\ref{d9b}) imply the NLO formula
\begin{align}
\run{h}^{-1} &\underset{\mathrm{NLO}}{=}
  \alpha_f^{-1} + \frac{1}{6\pi}(33-2f)\ln\frac{\bar{m}_h}{\bar{\mu}}
\nonumber \\
&\phantom{\underset{\mathrm{NLO}}{=}}
  + \frac{153-19f}{2\pi(33\!-\!2f)}
  \ln\Bigl[1 + \frac{\alpha_f}{6\pi}(33-2f)\ln\frac{m_h}{\bar{\mu}}\Bigr] .
\label{y2}
\end{align}
Again, Eq.~(\ref{y1}) can be used to write $\ln(\bar{m}_h/\bar{\mu})$ 
in terms of $\ln(m_h/\bar{\mu})$.  This leads to the following 
asymptotic formula for $\run{h}$ as $m_h \to \infty$:
\begin{gather}
\run{h}\ \sim\ 6\pi \!\Bigm/\! \Bigl\{
    (33-2f)\ln\frac{m_h}{\bar{\mu}} 
  + K_f\ln\ln\frac{m_h}{\bar{\mu}} + O(1)\Bigr\}  
 , \nonumber \\
K_f = \frac{3(153-19f)}{33-2f} - \frac{12(33-2f)}{31-2f} .
\end{gather}
These results show that we have complete control over the matching process.
Once closed expressions such as (\ref{y3}) and (\ref{y2}) have been 
obtained, RG invariance can be maintained for each logarithmic order, 
and so there is no need to truncate to a given order of perturbation 
theory.

\section{Mass-matching function}\label{mmf}

Most of the analysis above is restricted to the case of just one heavy 
quark $h$, but it can be readily extended to include \emph{sequential} 
decoupling, where heavy quarks are decoupled \emph{one at a time}.  
The new feature which arises is the need to match the mass of the 
second heavy quark.  For example, suppose that, having decoupled the 
$t$ quark in $F=6$ flavor QCD, we would like to decouple the $b$ quark 
as well:
\begin{equation} 
m_t \to \infty\ \ \mbox{first},\ \mbox{then}\ \ m_b \to \infty.
\label{j8}
\end{equation}
Then it will be necessary to match the six-flavor definition 
${m_b}_6 = m_b$ of the bottom quark mass to its five-flavour 
definition ${m_b}_5$.

As for the matching of couplings, the key is to find a RG invariant 
definition of mass to which the Appelquist-Carrazone theorem \cite{AC} 
can be applied.  This problem was solved by Bernreuther \cite{B1}, 
again by recourse to the momentum subtraction (MO) scheme.

Let $A_\ell(p^2)$ and $B_\ell(p^2)$ denote the form factors
for the unrenormalized 1PI light-quark self-energy amplitude\, 
$-i(/\hsp{-2}p A_\ell - {m_0}_\ell B_\ell)$.  This corresponds 
to the unrenormalized quark propagator
\begin{eqnarray}
S(p)
&=& \frac{i}{/\hsp{-2}p(1\!-\!A_\ell) 
                           - {m_0}_\ell(1\!-\!B_\ell)}
\nonumber \\
&=& \bigl(1\!-\!A_\ell\bigr)^{-1}
  \frac{i}{/\hsp{-2}p - {m_0}_\ell(1\!-\!B_\ell)/(1\!-\!A_\ell)}.
\end{eqnarray}
Define MO light-quark masses $M^\mathrm{MO}_\ell(Q)$ at a fixed 
space-like point $p^2 = -Q^2$:
\begin{equation}
M^\mathrm{MO}_\ell(Q) 
= {m_0}_\ell\bigl(1\!-\!B_\ell(-Q^2)\bigr)\!\bigm/\!
                            \bigl(1\!-\!A_\ell(-Q^2)\bigr) .
\end{equation}
This mass is RG invariant because, expressed in terms of 
renormalized quantities, it is finite.  The choice of space-like 
subtraction point $-Q^2$ means that mass renormalization, as well as 
coupling-constant and wave-function renormalization, is performed 
off-shell:%
\footnote{A similar MO definition for heavy-quark masses $M_h$ 
\cite{Appelquist} yields a $\beta$-function $\beta(g,M_h/Q)$ 
\cite{RGGP} with smooth threshold behavior at $Q \sim M_h$.}
\begin{equation} iS^{-1}(p)\bigr|_{p^2 = -Q^2}
  = /\hsp{-2}p - M^\mathrm{MO}_\ell(Q) 
  \not= 0 \ \mbox{at}\  p^2 = -Q^2.
\end{equation}
This avoids problems with Bloch-Nordsieck logarithms 
produced by $n$-loop perturbation theory at the on-shell 
point $p^2 \sim m^2$ \cite{AC}:
\[ iS^{-1}(p) \sim (p^2\hsp{-0.2}-\hsp{-0.2}m^2)^{-1}
 \bigl[\ln(p^2\hsp{-0.2}-\hsp{-0.2}m^2)\bigr]^n .  \]

A complication familiar to many authors \cite{gauge-dep} is that, 
unlike an on-shell renormalized mass, $M^\mathrm{MO}_\ell(Q)$\, is 
gauge dependent.  Despite this, the resulting mass-matching relation 
between $m_{\ell f}$ and $m_{\ell(f+1)}$ is gauge invariant \cite{B1} 
because the relations between $\overline{\mbox{\small MS}}$ masses and 
their bare counterparts are gauge invariant.  In two-loop perturbation 
theory, the result \cite{B1} for QCD with three colors and $f$ light 
flavours is
\begin{align}
\frac{m_{\ell f}}{m_{\ell(f+1)}} =
  1 &+ \frac{\alpha_{f+1}^2}{12\pi^2} 
      \biggl(\ln^2 \frac{m_h^2}{\bar{\mu}^2} + 
        \frac{5}{3} \ln\frac{m_h^2}{\bar{\mu}^2} +
        \frac{89}{36} \biggr)    \nonumber \\
    &+ O\bigl(\alpha_{f+1}^3\bigr) . 
\label{g0a}  
\end{align}

We would like to extend this result to include all terms of the same
\emph{logarithmic} order.  This is achieved by introducing our second 
matching function, ${\cal G}$ -- the matching function for mass 
renormalization.  On order to reduce notational complexity, we
consider the special case $f=5$ mentioned at the beginning of this
section.

Consider the RG equation
\begin{equation}
{\cal D}_6 \ln\frac{{m_b}_6}{{m_b}_5}\;
=\; \delta_6(\alpha_6) - \delta_5(\alpha_5)   
\label{g1d} 
\end{equation}
The leading power in a large-$m_t$ expansion of\, ${m_b}_6/{m_b}_5$\, 
is a function of\, $\alpha_5$ and\, $\ln(m_t/\bar{\mu})$\, but does not
depend on light-quark masses, so the general solution of (\ref{g1d}) is
\begin{eqnarray}
\ln\frac{{m_b}_6}{{m_b}_5}
&=& \int_{\alpha_5}^{\run{t}}\hsp{-2}dx\,
                   \frac{\delta_5(x)}{\beta_5(x)}
   - \int_{\alpha_6}^{\run{t}}\hsp{-2}dx\,
                   \frac{\delta_6(x)}{\beta_6(x)}\,
\nonumber \\[1mm]
&\phantom{=}&
   +\ {\cal G}_{6\to 5}(\run{t})\, +\, O(m_t^{-1}) 
\label{g2} 
\end{eqnarray}
where ${\cal G}_{6\to 5}(\run{t})$ is the \textsl{mass-matching function}. 
Like\, ${\cal F}_{6\to 5}(\widetilde{\alpha}_t)$\, in Eq.~(\ref{c6}), 
${\cal G}_{6\to 5}$ arises as an integration constant of a
RG equation, so it can depend only on the RG invariant\, $\run{t}$.
Also like ${\cal F}$, it cannot be deduced from the RG and must be 
calculated separately.

\begin{figure}[b] 
\begin{center}
\setlength{\unitlength}{1.5mm}
\begin{picture}(28.6,6.88)(0,-4.3)
\linethickness{0.25mm}
\newcommand{\curveglue}%
{\curve(0,0, 0.1,0.7, 0.8,0.8)\curve(0.8,0.8, 1.4,0.9, 1.7,1.46)%
\curve(1.7,1.46, 2.05,2, 2.65,2)\curve(2.65,2, 3.28,1.89, 3.75,2.3)%
\curve(3.75,2.3, 4.3,2.58, 4.85,2.3)%
\curve(4.85,2.3, 5.32,1.89, 5.95,2)\curve(5.95,2, 6.55,2, 6.9,1.46)%
\curve(6.9,1.46, 7.2,0.9, 7.8,0.8)\curve(7.8,0.8, 8.5,0.7, 8.6,0)}
\put(10,0){\curveglue}
\put(10,0){\curve(0,0, 4.3,-2, 8.6,0)}
\curve(2,0, 10,0)
\curve(18.6,0, 26.6,0)
\put(10,0){\circle*{0.5}}
\put(18.6,0){\circle*{0.5}}
\put(6,0.7){\makebox[0mm]{\small $b$}}
\put(22.6,0.7){\makebox[0mm]{\small $b$}}
\put(14.3,-4.3){\makebox[0mm]{\small $b$}}
\end{picture}
{\caption[Delta]
{One-loop correction to mass renormalization.} \label{fig1}} 
\end{center} 
\end{figure}
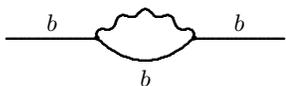

At one-loop order, there are no top-quark corrections (Fig.~\ref{fig1}),
so the mass-matching condition is trivial \cite{B1}: 
\begin{eqnarray} 
M_b^\mathrm{MO}
&=& 
m_{b,6} \bigl[1 + \alpha_6\{\mbox{1-loop}\}\bigr]\,
                         + O(\alpha_6^2)\,
\nonumber \\
&=&  
m_{b,5} \bigl[1 + \alpha_5\{\mbox{1-loop}\}\bigr]\,
                         +\, O(\alpha_5^2) .
\end{eqnarray}
Here $\alpha_f\{\mbox{1-loop}\}$ denotes the self-energy amplitude
derived from Fig.~\ref{fig1}.  Eliminating\, $M_b^\mathrm{MO}$, we find
\begin{equation}
\ln\frac{{m_b}_6}{{m_b}_5}\, 
 =\, (\alpha_5 - \alpha_6)\{\mbox{1-loop}\}\, +\, O(\alpha_5^2)\,
 =\,O(\alpha_5^2) . 
\label{g4} 
\end{equation}

At this point, we need to specify what is LO, NLO, and so on.  If we were
talking only about corrections to mass, we might consider terms\,
$\sim \bigl(\alpha_f\ln m_t/\bar{\mu}\bigr)^n$\, as LO, but in general,
it is more convenient to regard them as NLO.  That is because mass
renormalization does not contribute to physical amplitudes in LO.
With this convention, Eqs.~(\ref{d2}), (\ref{d9}), and (\ref{g1}) imply
\begin{align}
\int_{\alpha_f}^{\run{t}}\hsp{-2}dx\,
               \frac{\delta_f(x)}{\beta_f(x)}\;
&\underset{\mbox{\scriptsize NLO}}{=}\; 
      \frac{6}{\frac{33}{2}\!-\!f}
      \ln\frac{\run{t}}{\alpha_f}
\nonumber \\
&\underset{\mbox{\scriptsize NLO}}{=}\; 
      -\,\frac{6}{\frac{33}{2}\!-\!f}
  \ln\biggl[ 1 + \frac{\alpha_f}{3\pi}\biggl(\frac{33}{2}-f\biggr)
         \ln\frac{m_t}{\bar{\mu}}\biggr] .
\label{g5} 
\end{align}
In lowest-order perturbation theory, we have
\begin{align}
\int_{\alpha_5}^{\run{t}}\hsp{-2}dx\,
              &\frac{\delta_5(x)}{\beta_5(x)}
 - \int_{\alpha_6}^{\run{t}}\hsp{-2}dx\,
               \frac{\delta_6(x)}{\beta_6(x)}\,
\nonumber \\
&=
\frac{2}{\pi}\bigl(\alpha_6 - \alpha_5\bigr)
       \ln\frac{m_t}{\bar{\mu}} + O\bigl(\alpha_5^2\bigr)
= O\bigl(\alpha_5^2\bigr)
\label{g6} 
\end{align}
since the couplings $\alpha_6$ and $\alpha_5$ differ by $O(\alpha_5^2)$.
{}From Eqs.~(\ref{g2}), (\ref{g4}) and (\ref{g6}), we conclude
\begin{equation}
{\cal G}_{6\to 5}(\run{t})\, =\, O(\run{t}^2) .
\end{equation}

Having established that ${\cal G}$ is irrelevant at NLO, we neglect it
in Eq.~(\ref{g2}) and substitute (\ref{g5}). This yields the complete 
NLO expression:
\begin{align}
\ln\frac{{m_b}_6}{{m_b}_5}\;
&\underset{\mbox{\scriptsize NLO}}{=}\;
 \frac{6}{\frac{33}{2}\!-\!6}
  \ln\biggl[ 1 + \frac{\alpha_6}{3\pi}\biggl(\frac{33}{2}-6\biggr)
         \ln\frac{m_t}{\bar{\mu}}\biggr]
\nonumber \\
&\phantom{\underset{\mbox{\scriptsize NLO}}{=}}\;
  - \frac{6}{\frac{33}{2}\!-\!5}
  \ln\biggl[ 1 + \frac{\alpha_5}{3\pi}\biggl(\frac{33}{2}-5\biggr)
         \ln\frac{m_t}{\bar{\mu}}\biggr] . 
\label{g8} 
\end{align}
If desired, $\alpha_6$ can be eliminated in favour of $\alpha_5$ via
Eq.~(\ref{a8}).  Note that the $O(\alpha_5^2)$ NLO term is a double 
logarithm which arises from the diagram with a $t$-loop inserted in the 
gluon propagator (Fig.~\ref{fig2}):
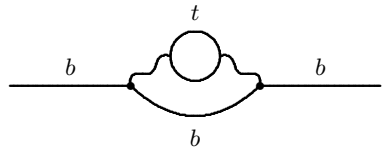
\begin{figure}[b] 
\begin{center}
\setlength{\unitlength}{2mm}
\begin{picture}(28.6,10)(0,-4)
\linethickness{0.25mm}
\newcommand{\curveglue}%
{\curve(0,0, 0.1,0.7, 0.8,0.8)\curve(0.8,0.8, 1.4,0.9, 1.7,1.46)%
\curve(1.7,1.46, 2.05,2, 2.65,2)%
\curve(5.95,2, 6.55,2, 6.9,1.46)%
\curve(6.9,1.46, 7.2,0.9, 7.8,0.8)\curve(7.8,0.8, 8.5,0.7, 8.6,0)}
\put(14.3,2){\bigcircle{3.3}}
\put(10,0){\curveglue}
\put(10,0){\curve(0,0, 4.3,-2, 8.6,0)}
\curve(2,0, 10,0)
\curve(18.6,0, 26.6,0)
\put(10,0){\circle*{0.5}}
\put(18.6,0){\circle*{0.5}}
\put(6,0.7){\makebox[0mm]{\small $b$}}
\put(22.6,0.7){\makebox[0mm]{\small $b$}}
\put(14.3,-4){\makebox[0mm]{\small $b$}}
\put(14.3,4.4){\makebox[0mm]{\small $t$}}
\end{picture}
{\caption[Delta]
{Two-loop heavy-quark contribution to mass renormalization of the bottom
quark.} \label{fig2}} 
\end{center} 
\end{figure}
\begin{align}
\ln\frac{{m_b}_6}{{m_b}_5}\;
&\underset{\mbox{\scriptsize NLO}}{=}\; 
 \frac{2}{\pi}\bigl(\alpha_6 - \alpha_5\bigr)
       \ln\frac{m_t}{\bar{\mu}}
 - \frac{\alpha_6^2}{3\pi^2}\biggl(\frac{33}{2}\!-\!6\biggr)
       \ln^2\frac{m_t}{\bar{\mu}}
\nonumber \\
&\phantom{\underset{\mbox{\scriptsize NLO}}{=}}\;\
 + \frac{\alpha_5^2}{3\pi^2}\biggl(\frac{33}{2}\!-\!5\biggr)
       \ln^2\frac{m_t}{\bar{\mu}} + O\bigl(\alpha_5^3\bigr) 
\nonumber \\
&\underset{\mbox{\scriptsize NLO}}{=}\;
 -\ \frac{1}{3}\biggl(\frac{\alpha_5}{\pi}\biggr)^2
       \ln^2\frac{m_t}{\bar{\mu}} + O\bigl(\alpha_5^3\bigr). 
\label{g9} 
\end{align}
This reproduces the NLO term of Bernreuther's result (\ref{g0a}) for%
\footnote{
Note that Bernreuther expands in $\alpha_6$ instead of $\alpha_5$, 
but to this order the coefficient is the same.}
$m_f/m_{f+1}$.

Eq.~(\ref{g8}) generates the complete set of NLO logarithms.  They 
correspond to diagrams with a string of one-loop $t$-quark bubbles 
inserted in the gluon propagator of the one-loop $b$-quark self-energy 
amplitude.

The constant term $89\alpha_f^2/(432\pi^2)$ in Eq.~(\ref{g0a})
corresponds to the first non-zero contribution to the matching 
function ${\cal G}$.  We state the result for any value of $f$:
\begin{equation}
{\cal G}_{f+1 \to f}(\run{h}) 
 = - \frac{89}{432\pi^2}\run{h}^2 + O(\run{h}^3).
\label{h2} 
\end{equation}
This term is required if NNNLO corrections are being calculated.

\section{APPLICATION TO SEQUENTIAL DECOUPLING}\label{asd}

When decoupling the $b$ quark, it is natural to define five-flavor
quantities $\mbox{$\run{b}$}_5$ and $\mbox{$\bar{m}_b$}_5$ 
by analogy with the six-flavor running coupling $\run{t}$ 
and mass $\bar{m}_t$ for the top quark: \vspace{-1mm}
\begin{eqnarray}
\ln\frac{{m_b}_5}{\bar{\mu}}
&=& \int_{\alpha_5}^{\mbox{$\scriptstyle \run{b}$}_5}
      \hsp{-1}dx\,\frac{1-\delta_5(x)}{\beta_5(x)}
\label{h3a} \\[0.5mm]
\ln\frac{\mbox{$\bar{m}_b$}_5}{{m_b}_5}
&=& \int_{\alpha_5}^{\mbox{$\scriptstyle \run{b}$}_5}
      \hsp{-1}dx\,\frac{\delta_5(x)}{\beta_5(x)} .
\label{h3}
\end{eqnarray}
Clearly, both $\mbox{$\run{b}$}_5$ and $\mbox{$\bar{m}_b$}_5$ 
are RG$_{f=5}$ invariant,
\begin{equation}
{\cal D}_5\mbox{$\run{b}$}_5 = 0
\hsp{4},\hsp{4}
{\cal D}_5\mbox{$\bar{m}_b$}_5 = 0
\end{equation}
but in fact, they are also RG$_{f=6}$ invariant as a consequence of
Eqs.~(\ref{e1}) and (\ref{eq:e2}):
\begin{equation}
{\cal D}_6\mbox{$\run{b}$}_5 = 0
\hsp{4},\hsp{4}
{\cal D}_6\mbox{$\bar{m}_b$}_5 = 0 .
\end{equation}

This means that $\mbox{$\widetilde{\alpha}_b$}_5$ and 
$\mbox{$\bar{m}_b$}_5$ can be expressed in terms of invariants
of the original six-flavor theory.%
\footnote{  
This property is essential for any generalization of the analysis 
to simultaneous decoupling, where all couplings and masses, running 
or otherwise, will have to be defined \emph{only} in terms of 
the initial theory ($F=6$) or the residual theory 
($f=4$ if just the $t$ and $b$ are being decoupled), with 
no reference to five-flavor couplings or masses.}
To see this, first combine Eqs.~(\ref{h3}) and (\ref{c5}), and 
then (\ref{g2}):
\begin{align}
\ln\frac{\bar{m}_t}{m_t}
-\ln\frac{\mbox{$\bar{m}_b$}_5}{{m_b}_5}
&= \int_{\alpha_6}^{\run{t}}
      \hsp{-1}dx\,\frac{\delta_6(x)}{\beta_6(x)}
   -\int_{\alpha_5}^{\mbox{$\scriptstyle \run{b}$}_5}
      \hsp{-1}dx\,\frac{\delta_5(x)}{\beta_5(x)}
\nonumber \\[0.5mm]
&= - \ln\frac{{m_b}_6}{{m_b}_5} 
 + \int_{\mbox{$\scriptstyle \run{b}$}_5}^{\run{t}}
      \hsp{-1}dx\,\frac{\delta_5(x)}{\beta_5(x)}
\nonumber \\[0.5mm]
&\phantom{+}\
 + {\cal G}_{6\to 5}(\run{t}) + O(m_t^{-1}) .
\end{align}
This equation simplifies to
\begin{align}
\ln\frac{\bar{m}_t}{\mbox{$\bar{m}_b$}_5}\,
&=\, \ln\biggl(\frac{m_t}{m_b}\biggr)_{\!6} 
  + \int_{\mbox{$\scriptstyle \run{b}$}_5}^{\run{t}}
      \hsp{-1}dx\,\frac{\delta_5(x)}{\beta_5(x)}
\nonumber \\
&\phantom{=}\ + {\cal G}_{6\to 5}(\run{t}) + O(m_t^{-1}) .
\label{h7} 
\end{align}
The sum of Eqs.~(\ref{h3a}) and (\ref{h3})
\begin{equation}
\ln\frac{\mbox{$\bar{m}_b$}_5}{\bar{\mu}}
= \int_{\alpha_5}^{\mbox{$\scriptstyle \run{b}$}_5}
     \hsp{-1}dx\,\frac{1}{\beta_5(x)}  .
\end{equation}
can be subtracted from Eq.~(\ref{c9}), with the result
\begin{equation}
\ln\frac{\bar{m}_t}{\mbox{$\bar{m}_b$}_5} = 
\int_{\mbox{$\scriptstyle \run{b}$}_5}^{\run{t}}
\hsp{-1}dx\, \frac{1}{\beta_5(x)} 
+ {\cal F}_{6\to 5}(\run{t})
+ O(m_t^{-1}) .
\label{i0} \end{equation}
If (\ref{i0}) is now combined with (\ref{h7}), we find that 
$\mbox{$\run{b}$}_5$ and hence $\mbox{$\bar{m}_b$}_5$
can be expressed in terms of RG invariants of the original six-flavor
theory, \emph{viz.}~$\run{t}$ and the ratio $(m_t/m_b)_6$:
\begin{align}
\ln\biggl(\frac{m_t}{m_b}\biggr)_{\!6}\, 
&=
\int_{\mbox{$\scriptstyle \run{b}$}_5}^{\run{t}}
\hsp{-1}dx\, \frac{1-\delta_5(x)}{\beta_5(x)} 
\nonumber \\
&\phantom{=}\
+ {\cal F}_{6\to 5}(\widetilde{\alpha}_t) 
- {\cal G}_{6\to 5}(\widetilde{\alpha}_t)\, 
+\, O(m_t^{-1}) .
\label{i1} \end{align}

The sequential decoupling of the $t$ and $b$ quarks refers to the
limiting procedure
\begin{equation}
\ln(m_t/\bar{\mu}) \gg \ln(m_b/\bar{\mu}) \rightarrow \infty 
\end{equation}
where we choose a six-flavor definition for $m_b$ as well as $m_t$. 
Leading-power six-flavor amplitudes are represented by logarithmic 
expansions for $t$-quark decoupling
\begin{equation}
\widetilde{\cal A}_6 \sim \sum_{p \geqslant 0} \widetilde{\cal C}_{p5} 
        \ln^p\bigl(m_t/\bar{\mu}\bigr) 
\end{equation} 
where each five-flavor coefficient $\widetilde{\cal C}_{p5}$ is a 
leading-power asymptotic expansion for $b$-quark decoupling:
\begin{equation}
\widetilde{\cal C}_{p5} \sim \sum_{q \geqslant 0} {\cal C}_{pq4} 
        \ln^q\bigl(m_b/\bar{\mu}\bigr) 
\end{equation} 
The last decoupling (that of the $b$ quark) is carried out with 
$\alpha_4$ held fixed.  Therefore we seek formulas for couplings 
such as $\alpha_6$ and $\run{t}$ in terms of $m_t$, $m_b$ and $\alpha_4$.

As always, the key step in the derivation of NLO formulas is the neglect
of some matching functions.  In this case, we neglect the NNLO functions
${\cal F}_{6 \to 5} = O(\run{t})$ and 
${\cal F}_{5\to 4} = O(\mbox{$\run{b}$}_5)$
for matching $\alpha_6$ to $\alpha_5$ and  $\alpha_5$ to  $\alpha_4$,
and the NNNLO function ${\cal G}_{6 \to 5} = O(\run{t}^2)$ for 
$\mbox{$m_b$}_5$  to be matched to $\mbox{$m_b$}_6 = m_b$.

We start with the NLO formula (\ref{g8}) for the five-flavor 
mass $\mbox{$m_b$}_5$.  To this order, all dependence on $\alpha_5$ 
and $\alpha_6$ can be eliminated via LO formulas derived 
from Eq.~(\ref{a8}),
\begin{align}
\alpha_5\, &\underset{\mbox{\scriptsize LO}}{=}\, \alpha_4\!\Bigm/\!\Bigl(1 
            + \frac{\alpha_4}{3\pi}\ln\frac{m_b}{\bar{\mu}}\Bigr)
\label{p0}\\
\alpha_6\, &\underset{\mbox{\scriptsize LO}}{=}\, \alpha_4\!\Bigm/\!\Bigl\{1 
            + \frac{\alpha_4}{3\pi}\Bigl(
   \ln\frac{m_t}{\bar{\mu}} + \ln\frac{m_b}{\bar{\mu}}\Bigr)\Bigr\}
\end{align}
where (again to this order) $\mbox{$m_b$}_5$ may be replaced by 
$m_b$ on the right-hand side. Then Eq.~(\ref{g8}) becomes
\begin{align}
\ln\frac{\mbox{$m_b$}_5}{\bar{\mu}}\; 
 &\underset{\mbox{\scriptsize NLO}}{=}\;
    \ln\frac{m_b}{\bar{\mu}} 
  - \frac{12}{23}\ln\Bigl(1 + 
        \frac{\alpha_4}{3\pi}\ln\frac{m_b}{\bar{\mu}}\Bigr)
\nonumber \\
 &\phantom{\underset{\mbox{\scriptsize NLO}}{=}}
  - \frac{8}{161}\ln\bigg(1 + \frac{23\alpha_4}{6\pi}\ln\frac{m_t}{\bar{\mu}}
       + \frac{\alpha_4}{3\pi}\ln\frac{m_b}{\bar{\mu}}\biggr)
\nonumber \\
 &\phantom{\underset{\mbox{\scriptsize NLO}}{=}}
  + \frac{4}{7}\ln\bigg(1 + \frac{\alpha_4}{3\pi}\ln\frac{m_b}{\bar{\mu}}
      + \frac{\alpha_4}{3\pi}\ln\frac{m_t}{\bar{\mu}}\biggr).
\label{mb5}
\end{align}

Similarly, consider the NLO relation (\ref{y3}) between
$\alpha_{f+1}$ and $\alpha_f$.  For the case $f=4$, the
heavy-quark mass $m_h$ in (\ref{y3}) is $\mbox{$m_b$}_5$, but we can use 
Eq.~(\ref{mb5}) to eliminate $\mbox{$m_b$}_5$ in favor of $m_b$,
\begin{align}
\alpha_5^{-1} \underset{\mathrm{NLO}}{=} \alpha_4^{-1} 
 &+ \frac{1}{3\pi}
      \ln\frac{\mbox{$m_b$}_5}{\bar{\mu}}\biggr|_\mathrm{NLO}\! 
  + c_4\ln\Bigl[1 + \frac{\alpha_4}{3\pi}\ln\frac{m_b}{\bar{\mu}}\Bigr]
\nonumber \\[1mm]
 &+ d_4\ln\Bigl[1 + \frac{25\alpha_4}{6\pi}
      \ln\frac{m_b}{\bar{\mu}}\Bigr] 
\label{p1}
\end{align}
where the constants $c_4$ and $d_4$ are given by
\begin{equation}
c_4 = 33/(23\pi)        
\hsp{4},\hsp{4}
d_4 = 121/(1150\pi). 
\end{equation}
For $f=5$, $m_h$ in (\ref{y3}) is the six-flavor mass $m_t$. Any
$\alpha_5$ dependence can be removed via Eqs.~(\ref{p0}) or (\ref{p1}):
\begin{align}
\alpha_6^{-1} \underset{\mathrm{NLO}}{=} \alpha_5^{-1}\Bigr|_\mathrm{NLO} 
 &+ \frac{1}{3\pi}\ln\frac{m_t}{\bar{\mu}}
    + c_5\ln\biggl[1 + \frac{\alpha_5}{3\pi}\Bigr|_\mathrm{LO}
           \ln\frac{m_t}{\bar{\mu}}\biggr]
\nonumber \\[1mm]
 &+ d_5\ln\biggl[1 + \frac{23\alpha_5}{6\pi}\Bigr|_\mathrm{LO}
      \ln\frac{m_t}{\bar{\mu}}\biggr] 
\label{p2}
\end{align}
The coefficients $c_5$ and $d_5$ have numerical values
\begin{equation}
c_5 = 47/(42\pi)        
\hsp{4},\hsp{4}
d_5 = 137/(966\pi). 
\end{equation}
The result of combining these formulas is:
\begin{align}
\alpha_6^{-1} &\underset{\mathrm{NLO}}{=} \alpha_4^{-1}
  + \frac{1}{3\pi}\Bigl(\ln\frac{m_b}{\bar{\mu}}
                        + \ln\frac{m_t}{\bar{\mu}}\Bigr)
\nonumber \\
&\phantom{\underset{\mathrm{NLO}}{=}}\
 + \frac{55}{42\pi}\ln\bigg(1 
      + \frac{\alpha_4}{3\pi}\ln\frac{m_b}{\bar{\mu}}
      + \frac{\alpha_4}{3\pi}\ln\frac{m_t}{\bar{\mu}}\biggr)
\nonumber \\
&\phantom{\underset{\mathrm{NLO}}{=}}\
 + \frac{121}{966\pi}\ln\biggl(1 
      + \frac{23\alpha_4}{6\pi}\ln\frac{m_t}{\bar{\mu}}
      + \frac{\alpha_4}{3\pi}\ln\frac{m_b}{\bar{\mu}}\biggr) 
\nonumber \\
&\phantom{\underset{\mathrm{NLO}}{=}}\
 + \frac{121}{1150\pi}\ln\biggl(1 
            + \frac{25\alpha_4}{6\pi}\ln\frac{m_b}{\bar{\mu}}\biggr).
\end{align}

The same procedure can be applied to the NLO formula (\ref{y1}) for 
the RG invariant mass $\bar{m}_h$. For $f=4$, the result is 
\begin{align}
\ln\frac{\mbox{$\bar{m}_b$}_5}{\bar{\mu}}
 &\underset{\mbox{\scriptsize NLO}}{=} \ln\frac{m_b}{\bar{\mu}} 
  - \frac{12}{23}\ln\Bigl(1 + 
        \frac{25\alpha_4}{6\pi}\ln\frac{m_b}{\bar{\mu}}\Bigr)
\nonumber \\
 &\phantom{\underset{\mbox{\scriptsize NLO}}{=}}
  - \frac{8}{161}\ln\bigg(1 + \frac{23\alpha_4}{6\pi}\ln\frac{m_t}{\bar{\mu}}
       + \frac{\alpha_4}{3\pi}\ln\frac{m_b}{\bar{\mu}}\biggr)
\nonumber \\
 &\phantom{\underset{\mbox{\scriptsize NLO}}{=}}
  + \frac{4}{7}\ln\bigg(1 + \frac{\alpha_4}{3\pi}\ln\frac{m_b}{\bar{\mu}}
      + \frac{\alpha_4}{3\pi}\ln\frac{m_t}{\bar{\mu}}\biggr)
\label{p4}
\end{align}
after substitution of Eq.~(\ref{mb5}) for $\mbox{$m_b$}_5$.
For $f=5$, Eq.\ (\ref{y1}) expresses $\bar{m}_t$ in terms of $\alpha_5$, 
for which the LO formula (\ref{p0}) may be substituted, with the result
\begin{align}
\ln\frac{\bar{m}_t}{\bar{\mu}}
&\underset{\mbox{\scriptsize NLO}}{=} \ln\frac{m_t}{\bar{\mu}}
  + \frac{4}{7}\ln\Bigl(1 
       + \frac{\alpha_4}{3\pi}\ln\frac{m_t}{\bar{\mu}}
       + \frac{\alpha_4}{3\pi}\ln\frac{m_b}{\bar{\mu}}\Bigr) 
\nonumber \\
&\phantom{\underset{\mbox{\scriptsize NLO}}{=}}\
   - \frac{4}{7}\ln\Bigl(1
       + \frac{23\alpha_4}{6\pi}\ln\frac{m_t}{\bar{\mu}}
       + \frac{\alpha_4}{3\pi}\ln\frac{m_b}{\bar{\mu}}\Bigr) .
\label{p5}
\end{align}
If desired, inverses of (\ref{p4}) and (\ref{p5}) can be constructed
and used to express quantities such as $\alpha_6$ in terms of the 
invariant masses $\mbox{$\bar{m}_b$}_5$ and $\bar{m}_t$ instead of 
$m_b$ and $m_t$.

Finally, we extract NLO formulas for the invariant running couplings
for sequential $t,b$ decoupling from Eq.~(\ref{y2}) for $\run{h}$.
For the case $f=4$, Eqs.~(\ref{y2}) and (\ref{p4}) imply
\begin{eqnarray}
\mbox{$\run{b}$}_5^{-1} 
 &\makebox[1em]{$\underset{\mathrm{NLO}}{=}$}& \alpha_4^{-1}
    + \frac{25}{6\pi}\ln\frac{m_b}{\bar{\mu}}
   - \frac{729}{1150\pi}\ln\biggl(1 
            + \frac{25\alpha_4}{6\pi}\ln\frac{m_b}{\bar{\mu}}\biggr)
\nonumber \\
&\phantom{=}& 
  -\, \frac{100}{483\pi}\ln\biggl(1 
      + \frac{23\alpha_4}{6\pi}\ln\frac{m_t}{\bar{\mu}}
      + \frac{\alpha_4}{3\pi}\ln\frac{m_b}{\bar{\mu}}\biggr) 
\nonumber \\
&\phantom{=}&
  +\, \frac{50}{21\pi}\ln\bigg(1 
      + \frac{\alpha_4}{3\pi}\ln\frac{m_b}{\bar{\mu}}
      + \frac{\alpha_4}{3\pi}\ln\frac{m_t}{\bar{\mu}}\biggr).
\end{eqnarray} 
For $f=5$, it is necessary to combine Eq.~(\ref{y2}) with 
Eqs.\ (\ref{mb5}), (\ref{p1}) and (\ref{p5}):
\begin{align}
\run{t}^{-1} &\underset{\mathrm{NLO}}{=} \alpha_4^{-1}
  + \frac{23}{6\pi}\ln\frac{m_t}{\bar{\mu}}
  + \frac{1}{3\pi}\ln\frac{m_b}{\bar{\mu}}     
\nonumber \\  
&\phantom{\underset{\mathrm{NLO}}{=}}\
 - \frac{457}{483\pi}\ln\biggl(1 
      + \frac{23\alpha_4}{6\pi}\ln\frac{m_t}{\bar{\mu}}
      + \frac{\alpha_4}{3\pi}\ln\frac{m_b}{\bar{\mu}}\biggr) 
\nonumber \\
&\phantom{\underset{\mathrm{NLO}}{=}}\
 + \frac{50}{21\pi}\ln\bigg(1 
      + \frac{\alpha_4}{3\pi}\ln\frac{m_b}{\bar{\mu}}
      + \frac{\alpha_4}{3\pi}\ln\frac{m_t}{\bar{\mu}}\biggr)
\nonumber \\
&\phantom{\underset{\mathrm{NLO}}{=}}\
 + \frac{121}{1150\pi}\ln\biggl(1 
            + \frac{25\alpha_4}{6\pi}\ln\frac{m_b}{\bar{\mu}}\biggr).
\end{align}  
A useful check of the formalism can be obtained by showing that the 
difference 
\[ \run{t}^{-1}-\mbox{$\run{b}$}_5^{-1}  \]
is correctly given by Eq.~(\ref{i1}) in NLO.

\section{Theories lacking Asymptotic Freedom}\label{nonAF}

So far, we have limited the discussion to heavy fermions in a gauge
theory such as QCD and used asymptotic freedom to obtain decoupling 
in the infinite-mass limit.

Does decoupling work if asymptotic freedom is not present, as in quantum 
electrodynamics (QED) or scalar field theory with $\lambda\phi^4$ 
interaction in four dimensions?   Perhaps this is part of the wider
debate \cite{trivial} about whether such theories are inconsistent or 
trivial, particularly for the continuum limit of the lattice approximation.

We start from the premise that a theory makes sense only if its heavy 
particles decouple in the infinite-mass limit to produce another 
consistent theory.  Questions about how the non-perturbative theory 
could depend on details of regulators and their removal are not 
considered.  We simply assume that some means of producing a fully
interacting cutoff-independent theory has been found, e.g.\ for QED$_k$ 
with $k$ species of equal-charge leptons, and apply our premise.

The notation is similar to that used above for QCD. Let $\alpha_k = 
e_k^2/(4\pi)$ be the $\overline{\mbox{MS}}$ renormalized fine structure
constant for QED$_k$, where the charged leptons have 
$\overline{\mbox{MS}}$ masses $m_j,\ j=1, \ldots ,k$, and let $\beta_k$
and $\delta_k$ be the Callan-Symanzik functions for charge and mass
renormalization.  Denote by $\run{H}$ and $\bar{m}_H$ Witten's invariant 
versions of the running fine-structure constant and heavy-lepton mass.
Then, repeating the arguments of Secs.~\ref{mf} and \ref{mmf}, we can 
construct coupling-constant and mass matching functions ${\cal F}$ and 
${\cal G}$ for the decoupling of one species of lepton:
\begin{equation}
\mbox{QED}_{k+1} \to \mbox{QED}_k,\  k \geqslant 1
\end{equation}
The free-photon theory QED$_0$ lacks a $\beta$-function so it is a
special case.

First consider the analogue of Eq.~(\ref{c8}), including the case $k=0$:
\begin{equation}
\ln\frac{\bar{m}_H}{\bar{\mu}}
= \int_{\alpha_{k+1}}^{\run{H}}\hsp{-2.5}dx\,\frac{1}{\beta_{k+1}(x)}
\end{equation}
The decoupling condition
\begin{equation}
\alpha_{k+1} \to 0\ \mbox{as}\ m_H \to \infty\ \mbox{for fixed}\ \alpha_k
\label{q1}\end{equation}
involves the $x=0$ solution of the equation $\beta_{k+1}(x) = 0$, but
it produces an infinity of the wrong sign because this fixed point is 
\emph{infra-red} stable.  For consistency, $\run{H}$ must approach a 
singularity of the integral sufficient to reverse the effect of the 
$x=0$ contribution. This could arise from an ultra-violet fixed point, 
\begin{equation}
\run{H} \to \alpha_{k+1,\infty}
\end{equation}
or else $\run{H}$ approaches $x=\infty$, in which case we must suppose that
$1/\beta_{k+1}(x)$ is not integrable.

Next define ${\cal F}$ as in Eq.~(\ref{c9}), with $k \geqslant 1$:
\begin{equation}
\ln\frac{\bar{m}_H}{\bar{\mu}}
= \int_{\alpha_k}^{\run{H}}\hsp{-2.5}dx\,\frac{1}{\beta_k(x)} 
       + {\cal F}_{k+1\to k}(\run{H}) 
\end{equation}
In the decoupling limit (\ref{q1}), the singularity on the left-hand side
is generated \emph{entirely} from the running of $\run{H}$.  We can
conclude that this is due to a QED$_k$ fixed point $\alpha_{k,\infty}$
\emph{only} if it coincides with that of QED$_{k+1}$:
\begin{equation}
\alpha_{k+1,\infty} = \alpha_{k,\infty}
\label{q2}\end{equation}
Similarly, a non-integrable singularity at $x=\infty$ can be the sole cause
only if this happens for both QED$_k$ and QED$_{k+1}$.  Otherwise, we must 
suppose that the matching function has a singularity at $\alpha_{k,\infty}$ 
or $\infty$.

The non-perturbative theory of QED of Johnson, Baker, Willey and 
Adler \cite{JBWA} is an example of the case (\ref{q2}).  Indeed, if the
arguments of Adler for an infinite-order zero at the fixed point are
applied to a many-species theory (all with the same charge), the result
is clear:  there is no dependence on the number of species.

Notice that these conclusions are driven by the lack of asymptotic freedom
of the initial theory.  For example, consider what happens in QCD when
a heavy quark decouples from the non-asymptotically free 17-flavor
theory to produce the 16-flavor theory with asymptotic freedom.  
Eq.~(\ref{c8}) for $F=17$ implies that $\run{h}$ increases. Thus in
Eq.~(\ref{c9}) with $f=16$, $\run{h}$ is driven towards a 
non-perturbative infra-red region of the residual theory.

\section{Concluding Remarks}\label{c}

The introduction of the matching functions ${\cal F}$ and ${\cal G}$ 
for coupling constants and masses (Eqs.~(\ref{c9}) and (\ref{g2}))
completes the theoretical structure needed for a systematic application 
of the RG to the decoupling of heavy particles.  We have considered just 
QCD and QED, but the field-theoretic principles are much the same for
any theory.  The main case still to be checked is a full RG analysis 
of the decoupling of heavy gauge bosons whose masses are induced by 
the Higgs mechanism.

In this article, we decoupled only one particle at a time (for simplicity)
and concentrated on field-theoretic aspects of the subject.  Actually, our 
work on matching functions arises from a need to consider the simultaneous
decoupling of more than one heavy quark in phenomenological applications.  
The challenge is to keep track of dependence on large logarithms such as 
\begin{equation} 
\ln\frac{m_t}{m_b} 
  = \ln\frac{m_t}{\bar{\mu}} - \ln\frac{m_b}{\bar{\mu}} 
\end{equation}
Since these logarithms do \emph{not} depend on the 
$\overline{\mbox{\small MS}}$ scale $\bar{\mu}$, the conventional tactic 
used in phenomenology for single heavy quarks fails: there is no way of 
making such logarithms small by choosing $\bar{\mu} = O(m_h)$.

As indicated in our work \cite{BCST} on NLO heavy-quark effects in 
axial charges of nucleons, the analysis can be generalized to include 
simultaneous decoupling of several heavy particles.  This includes the 
introduction of matching functions of several variables, one for each 
heavy particle.  We will present this extension of the theory in a 
forthcoming publication.

One can also anticipate generalizations to situations where momentum 
and mass logarithms grow large together.  Examples from the literature
occur in collider physics \cite{denner}, Higgs and supersymmetric 
particle production \cite{dawson}, and deep-inelastic scattering 
through thresholds for heavy-particle production \cite{blumlein}.

\begin{acknowledgments}
This work was supported by the Australian Research Council and the 
University of Adelaide.
S.D.B. thanks the Austrian Science Fund (FWF) for financial support 
(grant M683), and the CSSM for its hospitality at an important stage 
of this work.
F.M.S. is supported by contract number PV-IFT/005.
R.J.C. thanks Professor Wojtek Zakrzewski for his hospitality at Durham.
\end{acknowledgments}

\end{document}